\documentclass[journal,10pt]{IEEEtran}
\usepackage{amsthm}
\usepackage{amsfonts}
\usepackage{amssymb}
\usepackage{mathrsfs}
\usepackage{mathtools}
\usepackage{float}
\usepackage[pdftex]{graphicx}
\usepackage{amsmath}
\usepackage{color}
\usepackage{multirow}
\usepackage{graphicx}

\definecolor{color1}{RGB}{199,209,232}
\definecolor{color2}{RGB}{230,231,233}

\usepackage{verbatim}
\usepackage[table]{xcolor}
\usepackage{balance}
\usepackage{lipsum}
\usepackage[inline]{enumitem}
\usepackage{cuted}
\ifCLASSOPTIONcompsoc
\usepackage[caption=false,font=normalsize,labelfont=sf,textfont=sf]{subfig}
\else
\usepackage[caption=false,font=footnotesize]{subfig}
\fi
\usepackage{cite}
\usepackage{hyperref}
\usepackage{graphicx,float,wrapfig,epstopdf,amsmath}
\begin{document}
	
	\title{ Cognitive Learning-Aided Multi-Antenna Communications}
	
	\author{\IEEEauthorblockN{Ahmet M. Elbir \textit{Senior Member, IEEE} and Kumar Vijay Mishra \textit{Senior Member, IEEE}}
		\thanks{A. M. Elbir is with the SnT at the University of Luxembourg, Luxembourg and with Duzce University, Turkey (e-mail: ahmetmelbir@gmail.com).} 
		\thanks{K. V. Mishra is with the United States Army Research Laboratory, Adelphi, MD 20783 USA (e-mail: kumarvijay-mishra@uiowa.edu).}
	}
	
	\maketitle 
	
	\begin{abstract}
		Cognitive communications have emerged as a promising solution to enhance, adapt, and invent new tools and capabilities that transcend conventional wireless networks. Deep learning (DL) is critical in enabling essential features of cognitive systems because of its fast prediction performance, adaptive behavior, and model-free structure. These features are especially significant for multi-antenna wireless communications systems, which generate and handle massive data. Multiple antennas may provide multiplexing, diversity, or antenna gains that, respectively, improve the capacity, bit error rate, or the signal-to-interference-plus-noise ratio. In practice, multi-antenna cognitive communications encounter challenges in terms of data complexity and diversity, hardware complexity, and wireless channel dynamics. DL solutions such as federated learning, transfer learning and online learning, tackle these problems at various stages of communications processing, including multi-channel estimation, hybrid beamforming, user localization, and sparse array design. {\color{black}This article provides a synopsis of various DL-based methods to impart cognitive behavior to multi-antenna wireless communications for improved robustness and adaptation to the environmental changes while providing satisfactory spectral efficiency and computation times. We discuss DL design challenges from the perspective of data, learning, and transceiver architectures. In particular, we suggest quantized learning models, data/model parallelization, and distributed learning methods to address the aforementioned challenges. }
	\end{abstract}
	
	\begin{IEEEkeywords}
		Deep learning, Multi-antenna systems, Online learning, Federated learning. 
	\end{IEEEkeywords}
	\IEEEpeerreviewmaketitle
	
	
	\section{Introduction}
	As the wireless standards move from the recent initial deployment of fifth-generation (5G) new  radio to 6G, there is an urgency in transcending the current data rates, robustness, spectral reuse, latency, and energy  efficiency. {\color{black}Therefore, instead of a single antenna, nearly all recent wireless technologies employ multiple antennas to address these challenges~\cite{mimoOverviewGoeffreyLi}}. For example, the \textit{spatial diversity} provided by the use of multiple antennas is helpful in multipath scenarios. By providing multiple copies of the transmitted signal in statistically independent fading, this configuration reduces errors at the receiver. Further, \textit{spatial multiplexing} with multiple antennas allows transmission of many simultaneous data streams over a multipath channel and same frequency band, thereby improving bit rates. Finally, large antenna arrays offset the severe attenuation and path losses at higher frequencies through \textit{beamforming gain}.
	
	
	\textcolor{black}{Multiple antennas are able to offer \textit{more efficient} use of limited spectrum by \textit{also} including diversity in the spectral (multiple frequency allocations) and time (multiple delays of signal replica) domains.} In wireless communications, multiple-input single-output, multiple-input multiple-output (MIMO), and massive MIMO are common architectures~\cite{widebandChannelEst2}. The backscattered communications, which aid in passive sensing applications, also employ multiple antennas to detect and localize unknown emitters~\cite{dl_DOAEst}. 
	
\begin{figure*}[t]
		\centering
		{\includegraphics[draft=false,width=1.6\columnwidth]{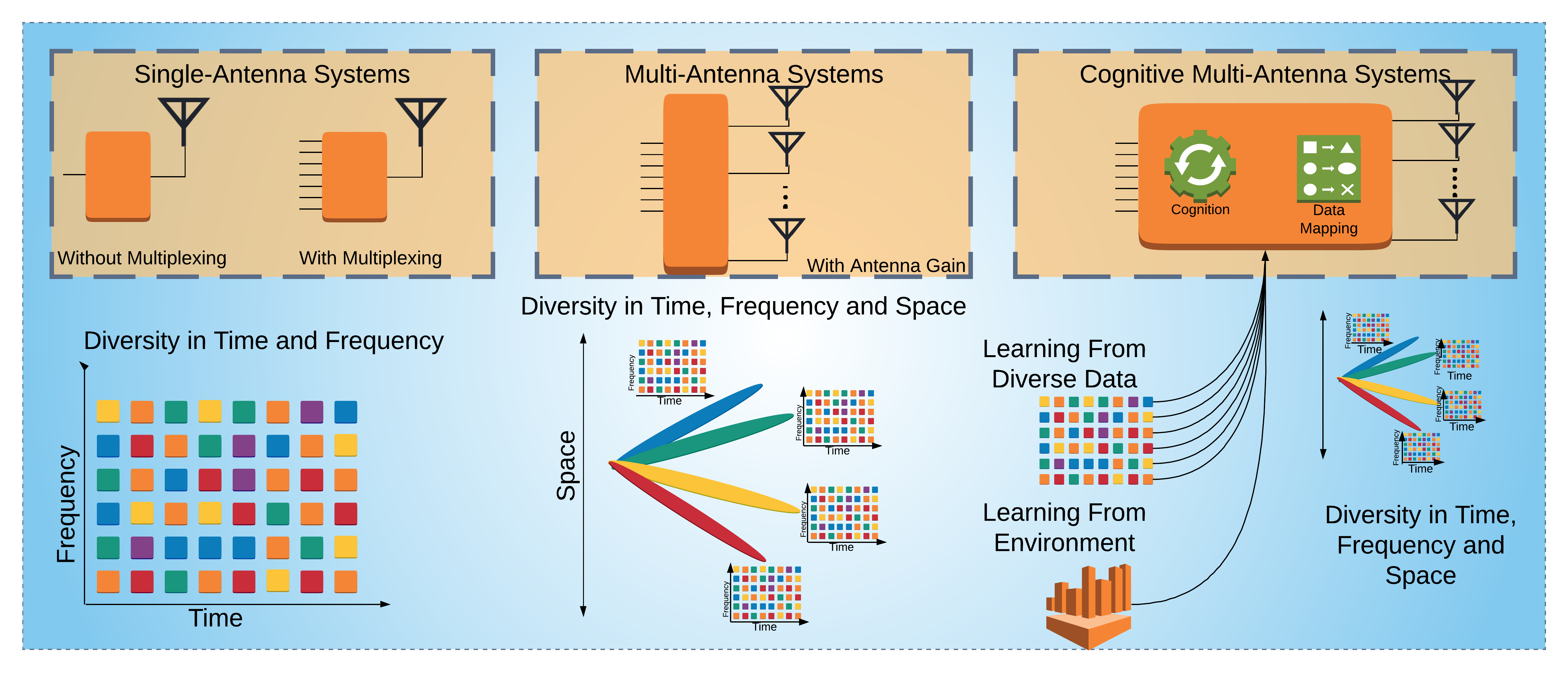} }
		\caption{Evolution of cognitive communications from single and multi-antenna systems to DL-based cognitive systems. Whereas single antenna provides multiplexing gain by jointly processing multiple data streams, the multi-antenna systems bring in beamforming gain and spatial diversity. The DL-based architectures learn from both diverse datasets and channel to enable multi-antenna systems with cognition.
		}
		\label{fig_CognitiveComm}
	\end{figure*}

	In recent years, multi-antenna technologies have increasingly leveraged cognitive processing and control (Fig.~\ref{fig_CognitiveComm}). These \textcolor{black}{cognitive systems are capable of quick and greater control of transmitters and higher adaptability of receivers than their non-cognitive counterparts.} The conventional optimization and analytical methods are not efficient or fast in adaptively utilizing the excess degrees of freedom and high-dimensional data in multi-antenna systems. This has motivated growing use of learning-based techniques to accomplish various stages of sense-learn-adapt cycle of cognitive multiple-antenna communications. The traditional learning-based methods, such as $k$-nearest neighbor algorithm, support vector	machines, $k$-means	clustering, principal and independent component
	analysis have been widely recognized for regression, classification and clustering tasks~\cite{ml_WCM}. These traditional methods have limited performance, especially for huge training data~\cite{dl_WCM}.	In this context, deep learning (DL) is a key cognition-enabling technology \cite{dl_GGui_WCM,dl_GeoffreyLi1} because it offers robustness against the imperfections in the data, fast prediction response, allows a model-free mapping between input and output data, \textcolor{black}{by training a learning model with a huge number of learnable parameters.} Recent research suggests promising outlook for DL in solving common wireless communications problems, such as signal detection~\cite{dl_GGui_WCM,dl_WCM}, channel estimation~\cite{elbir2019online} and beamforming~\cite{Vu2021Jan,elbir2019online}. 
	
	In this article, we focus on the developments in this area during the last three years, wherein novel DL architectures have emerged to enhance the adaptability of multi-antenna communications. Whereas traditional learning methods for possible use in wireless communications were discussed in \cite{ml_WCM}, simple multilayer perceptron (MLP) architectures for the physical layer design in 5G systems were examined in \cite{dl_GGui_WCM}. Recently, \cite{dl_WCM} concentrated on signal detection and decoding in DL-based transceivers for 5G non-orthogonal multiple access. Unlike these expositions, \textcolor{black}{the main contributions of this article are as follows:
		~\\
		${\bullet}$ We explore multi-antenna wireless applications of	existing and new DL architectures such as unsupervised learning (UL), supervised learning (SL), online learning (OL), reinforcement learning (RL), transfer learning (TL), and federated learning (FL). ~\\
		${\bullet}$ We  provide a comprehensive analysis for DL-based cognitive physical layer problems (Fig.~\ref{fig_DL_Applications}), such as signal detection, channel estimation, user localization, antenna selection and beamforming via different learning techniques, including some numerical examples.}
	
	\textcolor{black}{The rest of the paper is organized as follows. In the next section, we introduce the role of DL in combating the challenges in multi-antenna communications.	Section~\ref{sec:dl_sol} discusses DL-based solutions for multi-antenna communications. We follow this with unaddressed design challenges and consequent future research directions in Section~\ref{sec:chall}. We conclude in Section~\ref{sec:summ}. }

	\begin{table*}[t]
		\caption{Summary of Major Challenges in Multi-antenna Systems and DL-Based Solutions}
		\label{tableChallenges}
		\centering
		\begin{tabular}{p{0.21\textwidth} p{0.21\textwidth} p{0.21\textwidth}  p{0.21\textwidth} }
			\hline
			\hline
			\textbf{Challenge}\cellcolor{color1} & {\color{black}Multidimensional data \par and hardware  complexity} 
			
			\cellcolor{color2}  
			& {\color{black}MIMO channel dynamics \par  and interference}
			\cellcolor{color1} & {\color{black}Power and memory usage arising from multi-dimensional processing}
			\cellcolor{color2}  \\
			\textbf{DL advantage}\cellcolor{color2} & Fast post-training processing\cellcolor{color1} & Model-free architecture\cellcolor{color2} &Replacement of some hardware and processing \cellcolor{color1}  \\
			\hline
			\textbf{Enabling technology}\cellcolor{color1} & Parallel implementation\cellcolor{color2} & Huge number of learnable parameters\cellcolor{color1} & End-to-end learning\cellcolor{color2} \\
			\hline
			\color{black}\textbf{Related DL techniques and} \par \color{black}\textbf{multi-antenna applications} \cellcolor{color2} & \color{black}SL for HB, antenna selection~\cite{Vu2021Jan}, symbol detection~\cite{dl_WCM}, channel estimation~\cite{deepCNN_ChannelEstimation}; FL for HB~\cite{elbir2020FL} \cellcolor{color1} & \color{black}OL for HB and MIMO channel estimation~\cite{elbir2019online}; TL for antenna selection~\cite{elbir2020TL}; RL for HB~\cite{dl_GGui_WCM}\cellcolor{color2} & \color{black} FL for beamforming~\cite{elbir2020FL}; RL for detection and estimation~\cite{dl_GGui_WCM}; UL for resource allocation~\cite{unsupervisedMag} \cellcolor{color1} \\
			\hline
			\hline
		\end{tabular}
	\end{table*}
	
	\section{\color{black}Multi-Antenna Challenges and DL}
	The ever growing number of devices and applications implies greater data complexity and hardware challenges in wireless communications. {\color{black}These challenges scale with the number of antennas in multi-antenna systems. In the following, we investigate these challenges and introduce the DL methods with possible applications in multi-antenna systems.}
	

	\subsection{Challenges}
	Table~\ref{tableChallenges} summarizes the following key challenges in the prevalent multi-antenna wireless communications.
	\subsubsection{Data and hardware complexity} \textcolor{black}{For enhanced reliability, we desire low bit-error-rate (BER) transmissions with meta-data so that the quality of high-definition images/videos and voice data retain their quality.} This massive volume and diversity of data must be processed and transmitted in a few milliseconds to meet low latency requirements. Here, the model-free structure of DL {\color{black}is expected to greatly reduce the processing complexity~\cite{dl_WCM}. The time complexity of the DL methods comprises of data generation, model training, and prediction stages. While data generation and model training are time-consuming tasks, their computations constitute only one-time upfront cost. However, at the prediction stage, DL offers lower computation/time complexity than model-based signal processing techniques. This is made possible using parallel computations with graphical processing units (GPUs)~\cite{parallelizationRef} and also specialized GPUs, such as Intel Movidius that performs 2 trillion operations per second with 500 mW power consumption~\cite{mlAtTheEdge}.} Thus, to deal with the complex hardware requirements of state-of-the-art signal processing techniques, DL enjoys a model-free structure, which does not require diverse hardware components~\cite{Vu2021Jan,dl_GGui_WCM}. In fact, DL-based architectures require largely parallel computing power to construct a non-linear relationship between the input and output data.

	\subsubsection{Channel dynamics and interference} The 5G and 6G standards envisage greater use of higher frequency bands,	where the wireless channel has short coherence times~\cite{dl_WCM}. In particular, significant efforts have been undertaken to model the wireless channel characteristics and develop channel estimation techniques for the planned 5G deployment at over $30$ GHz~\cite{widebandChannelEst2}. In dynamic channels, such as indoor and vehicular scenarios, DL is helpful for achieving fast system reconfigurability, efficient feature extraction, and robust performance against the channel imperfections due to the rapid changes in  multipath, delay spread, angle spread and Doppler shift {\color{black}to improve the signal-to-interference-plus-noise ratio (SINR)}. The model-free structure and ability to learn the features hidden in the raw data~\cite{elbir2019online,dl_GGui_WCM} makes DL tolerant to interference from the coexisting emitter signals. \textcolor{black}{Furthermore, the temporal correlation of multiple antennas at the BS is exploited by the learning models with long short-term memory network for CSI prediction in multiple data frames.}
	
	\subsubsection{Power and memory} 
\begin{figure*}[t]
		\centering
		{\includegraphics[draft=false,width=1.8\columnwidth]{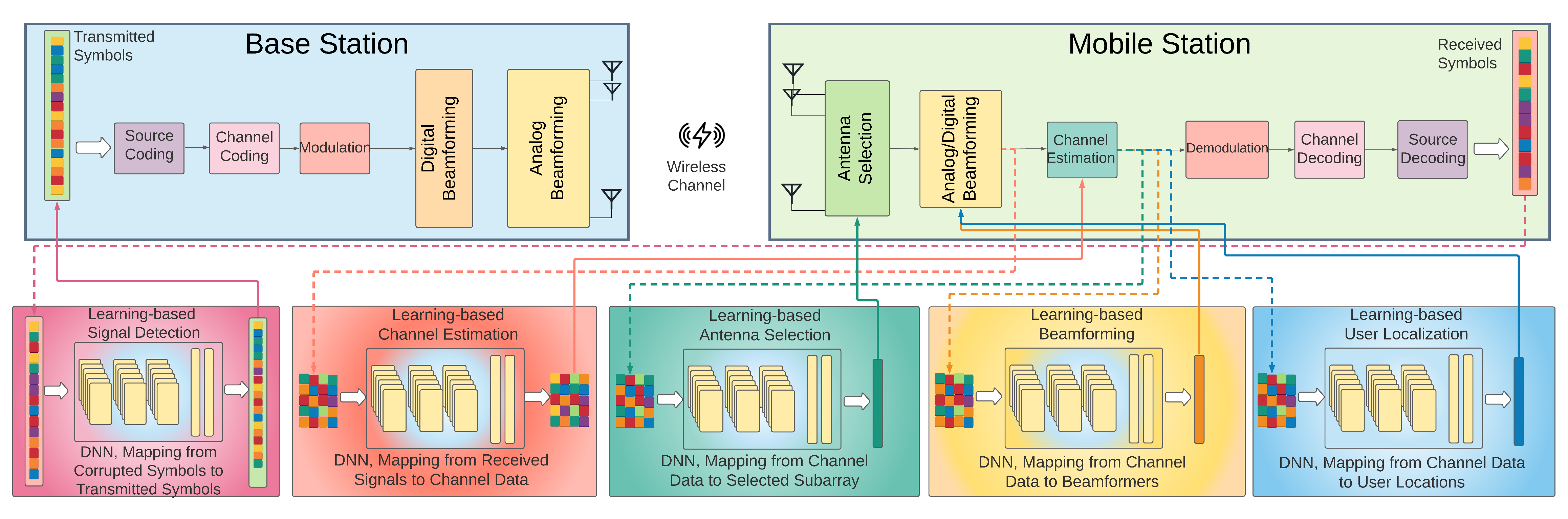} }
		\caption{A multi-antenna wireless communications scenario with DL architectures includes various cognitive communications applications, each of which has following different input-output pairs: signal detection (received symbols - transmitted symbols), channel estimation (raw received data - channel matrix), antenna selection (channel matrix - best subarray index/class), beamforming (channel matrix - beamformer weights) and user localization (channel covariance matrix - user DoA angles).
		}
		\label{fig_DL_Applications}
	\end{figure*}

The conventional wireless communication systems suffer from high power consumption. {\color{black}This only scales up with the large number of antennas in 5G systems.} Instead of processing data through several signal processing blocks (Fig.~\ref{fig_DL_Applications}), end-to-end DL techniques~\cite{dl_GGui_WCM,dl_WCM} provide a data mapping from the received and transmitted symbols directly and, as a result, considerably bring down power utilization. Since data caching/storing is not needed in the intermediate processing blocks, memory requirements also come down in such implementations~\cite{elbir2020FL}.

	\subsection{Deep Learning Techniques}
	Some of the more recent DL networks that we consider to address the aforementioned problems are as follows.
	{\color{black}\subsubsection{Unsupervised Learning} The UL identifies	hidden features/patterns in the ``untagged" data for which the pre-assigned labels or scores are not provided~\cite{unsupervisedMag}. This aids in wireless communications problems such as power (or resource) allocation, in which the objective function of the optimization problem is selected as the loss function of the learning model.}

	\subsubsection{Supervised Learning} Here, the labeled data are used for model training while minimizing the mean-squared-error (MSE) between the label and the model's response. \textcolor{black}{The SL is beneficial for tasks such as multi-channel estimation and beamforming, where labeled data could be employed from prior observations or data generation~\cite{deepCNN_ChannelEstimation,dl_GGui_WCM} (see Fig.~2).}
	
	\subsubsection{Online Learning} The performance of SL is subject to the similarity of the input data and the learned features in the training dataset. When the input data significantly deviate from that of training, the learning model is unable to recognize new incoming data resulting in  prediction/classification loss (see Fig.~\ref{fig_TL_OL}).  In OL, the learning model adapts to the new data by updating its parameters based on newly received data. {\color{black}The OL is useful in channel estimation and {\color{black}hybrid beamforming (HB)} for highly dynamic channels such as mmWave and THz~\cite{elbir2019online}, which invariably employ large arrays.}
	

	\subsubsection{Reinforcement Learning} The OL requires \emph{online datasets} to update the learning model, which entails a labeling step. In order to eliminate this expensive labeling, RL directly yields the output by optimizing the objective function of the learning model. {\color{black}For example, HB design problem is solved via RL without computing the labels for each instance of the training data. Instead, the beamformer weights are found through iterative updates of the model parameters by optimizing the spectral efficiency~\cite{dl_GGui_WCM}. }
	
	\subsubsection{Transfer Learning} In TL, a pre-trained model trained on a large dataset of a particular task is used to deliver the learned features to another task, where only a small dataset is available. \textcolor{black}{Compared to OL, TL also uses a pre-trained DL model but OL adapts to the same task while the latter learns a different task.} {\color{black}As an example, TL is used to map data from different antenna array geometries in~\cite{elbir2020TL}.}

	\subsubsection{Federated Learning} Instead of collecting the data from user devices at a cloud center, the FL computes model updates at the user level and aggregates them at the cloud server, thus leading to a very low communication overhead~\cite{elbir2020FL}. {\color{black}FL is privacy-preserving since it does not involve raw data transmission. While this is a major advantage in the applications, such as image/speech recognition~\cite{mlAtTheEdge}, the physical layer data, e.g., beamformers, channels, may not require privacy concerns. In~\cite{elbir2020FL}, FL is employed for  privacy-aware HB applications.}

	\section{DL-Based Solutions}
	\label{sec:dl_sol}
	The DL algorithms for wireless communications are trained with huge datasets to achieve a robust and reliable performance. The data diversity and multi-channel processing of multi-antenna communications complement this requirement. Fig.~\ref{fig_DL_Applications} illustrates various DL applications in the processing chain, such as signal detection, channel estimation, antenna selection, beamforming, and user localization.

	\subsection{Signal Detection}
	The signal detection via conventional wireless communication systems involves several blocks to process the data symbols, such as source coding, channel coding and modulation, as shown in Fig.~\ref{fig_DL_Applications}. To leverage DL for signal detection, \cite{dl_GGui_WCM} devised an MLP for mapping received data symbols to the transmit symbols, thereby constructing an end-to-end mapping from channel effected data and the transmitted true symbols. Once the MLP is trained on a dataset composed of received-transmitted data symbols, the users feed the learning model with the block of received symbols, which contain the imperfections due to the hardware impairments and the wireless channel, then the MLP yields the estimated transmitted symbols.	A major advantage of this approach is its simplicity that the learning model estimates the data symbols directly, without a prior stage for channel estimation. Thus, this method is helpful reducing the cost of channel acquisition.


	\subsection{Channel Estimation}
	Reliable channel estimation is critical to ascertain the effect of the channel on the transmitted signal and recovering the transmitted information. To this end, a DL model is trained with the raw received signal collected at the antenna array outputs. A common way to form the training data is to collect data during pilot training~\cite{deepCNN_ChannelEstimation,elbir2019online}, and the input-output pairs are selected as the data after beamforming and channel estimation blocks, as shown in Fig.~\ref{fig_DL_Applications}.	The DL network constructs a non-linear mapping between the antenna array data and the channel matrix (computed via analytical approaches offline) during the training stage. In \cite{deepCNN_ChannelEstimation}, channel estimation is performed by convolution-only neural networks. These have poorer estimation accuracy than the generalized convolutional neural networks (CNNs) whose final layers are fully connected \cite{elbir2019online}. While the convolutional layers are good at extracting the additional features inherent in the input, fully connected layers are more efficient in non-linearly mapping the input to the labeled data. The choice of the network layers is, therefore, important to guarantee high channel estimation accuracy. 
	
	\subsection{Sparse Array Design}
	It is highly desired to reduce the number of \textcolor{black}{active radio frequency chains} in large arrays used in massive MIMO communications. This has led to investigations in sparse arrays or using subarrays from a full array by selecting only a few elements. In general, searching for an optimal sparse antenna array is a combinatorial problem, whose computational complexity increases with the number of antennas. Since a closed-form solution is difficult to come by, several sub-optimal but mathematically tractable solutions have been proposed~\cite{dl_DOAEst}. Here, DL is helpful in decreasing the overhead of finding optimum sparse array.	 In \cite{Vu2021Jan}, DL chooses the best sparse array by evaluating the resulting system spectral efficiency as a performance metric. This is done by treating the problem as a classification task where it is intended to find the best sparse array among all subarray configurations. As illustrated in Fig.~\ref{fig_DL_Applications}, the classification model accepts the channel data and yields the best subarray index/class at the output. During model training, the sparse array design problem is solved offline. Then, the learned model is deployed to predict the optimal subarray in real-time, which significantly reduces computation time. In~\cite{Vu2021Jan}, only antenna selection at the receiver is studied. Being a more complex problem, the transmit antenna selection can be investigated for future research.

	\subsection{Hybrid Beamforming}
	In contrast to channel estimation and sparse array design, HB involves optimization of several high-dimensional variables at both transmitter and receiver. The DL-based HB 	significantly reduces the computational burden arising from the optimization of HB weights 	and hardware cost. The DL model input data is the channel matrix that is labeled by the corresponding beamformer weights~\cite{Vu2021Jan,dl_GGui_WCM} (Fig.~\ref{fig_DL_Applications}). In addition, raw array output data are used to directly obtain the HB weights, thus bypassing the channel estimation~\cite{elbir2019online}. In~\cite{Vu2021Jan} and \cite{elbir2019online}, CNNs are used for HB design whereas \cite{dl_GGui_WCM} employs an MLP architecture that consists of only fully connected layers. \textcolor{black}{Here, the superior performance of CNN over MLP is because of convolutional layers in the lower layers for enhanced feature extraction and fully connected layers in the higher layers for improved data mapping \cite{deepCNN_ChannelEstimation,elbir2019online}}.

	\subsubsection{Joint Hybrid Beamforming and Antenna Selection} It is possible to design twin CNNs to jointly solve HB design and antenna selection tasks~\cite{Vu2021Jan}. {\color{black}In this case, the best subarray and the HB weights are designed during training offline by maximizing the overall system's spectral efficiency or SINR}. Thus, two different datasets are generated for each task and two CNNs are trained accordingly. Once the learning models are trained, the channel data of the full array is fed to the first CNN (antenna selection) and the best subarray indices are obtained. Then, the channel matrix corresponding to the selected subarray is fed to the second CNN (HB), which yields the HB weights. The major advantage of this approach is to eliminate solving two problems separately. A single CNN may also be used but it may be less flexible to tune parameters of each individual problem. 
	
	\begin{figure*}[t]
		\centering
		{\includegraphics[draft=false,width=1.3\columnwidth]{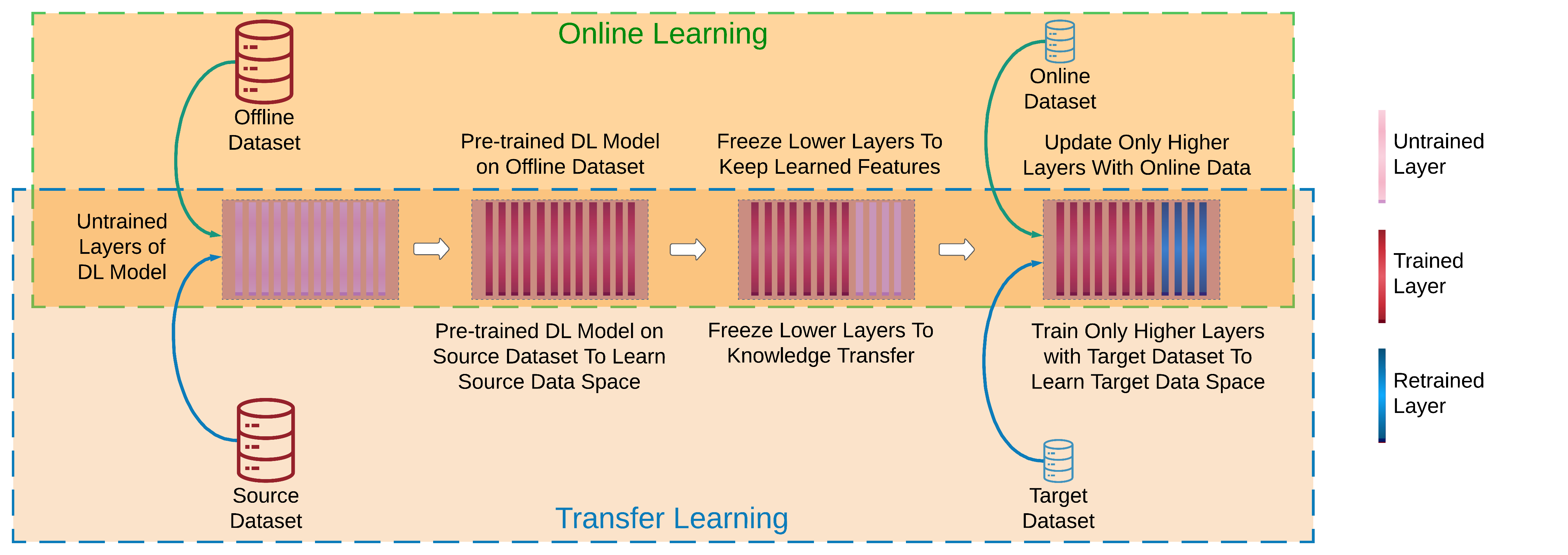} }
		\caption{The OL  framework (top) utilizes a pre-trained DL model trained on offline dataset and updates the model parameters with the online dataset. The TL framework (bottom) benefits from a pre-trained learning model that is trained on a large dataset (source), which differs from the small dataset (target) of the application under consideration, thus transferring the knowledge from the learned features to the target application.
		}
		\label{fig_TL_OL}
	\end{figure*}
	\subsubsection{Joint Hybrid Beamforming and Channel Estimation} A joint HB design with channel estimation is also possible with two different CNNs. In~\cite{elbir2019online}, different DL frameworks are proposed for various combinations of both tasks. These include HB design directly from the raw data; and channel estimation from the raw data, then HB design from the estimated channel for both narrowband and wideband systems. With a single CNN, the HB design for narrowband systems is simpler than the wideband scenario, which involves the complexity of large input data that is scalable with the number of subcarriers~\cite{deepCNN_ChannelEstimation}. While feeding the CNN with large data improves the feature extraction and representation performance, a reliable performance is still obtained by constructing a dedicated CNNs for each subcarrier, whose input size is smaller compared to the wideband case~\cite{elbir2019online}.

	\subsection{User Localization} 
	The spatial diversity of multi-antenna systems is helpful in direction finding (DF) for the user locations. Here, sparsity in the channel arising from few and clustered paths may be exploited for estimating the direction-of-arrival (DoA). Thus, the direction-aided algorithms can be helpful reducing the feedback communication overhead since instead of feedbacking the whole channel data, only fewer parameters, such as DoA angles and channel gains are transmitted. Then, the channel matrix can be reconstructed at the BS by using the feedback data.	Most DoA algorithms employ grid search whose complexity, especially in two dimensions (azimuth and elevation), is excessive.	
	The DL-based algorithm is helpful in reducing the search complexity as well as providing robustness against the imperfections, such as mutual coupling, gain/phase mismatch and multipath.	Here, the size of the training dataset (input data and the number of labels) increases exponentially with the number of sources~\cite{elbir2020TL}.
	


	\section{Design Challenges and Future Directions}
	\label{sec:chall}

	\subsection{Data-Related Challenges}
	{\color{black}The DL training dataset must extensively sample the data space in order to sufficiently represent the inherent data diversity of multi-antenna communications~\cite{mlAtTheEdge,dl_WCM}.} However, this is an arduous task because it requires the wireless devices to collect data under all possible channel conditions. As a result, there is always a looming risk of data mismatch in these solutions.
	
	\subsubsection{Insufficient Coverage of Data Space}
	When comprehensive training data are unavailable or insufficient, TL is employed to benefit from a pre-trained DL model that is trained on a large dataset (which is not directly derived from the application-at-hand). {\color{black}For example, in DL-based channel estimation~\cite{deepCNN_ChannelEstimation,dl_WCM,elbir2019online}, the data may be collected in sparse directions, and a learning model that is trained on a dense data can be re-trained with the sparse data for performance improvement.} Here, a pre-trained model developed for a certain task is reused as the starting point for a model on a different task (Fig.~\ref{fig_TL_OL}). The target model need not be trained afresh. Instead, portions of source model layers are kept ``frozen'' and the remaining model parameters are updated using new, smaller datasets~\cite{elbir2020TL}. \textcolor{black}{Then, the DL model is trained with a source scenario for which a huge dataset is readily available. This pre-trained model is reused for a different, data-insufficient target scenario.} In a cognitive communications scenario, a model is trained on a huge dataset involving extensive channel realizations. Then, a TL method redeploys the same model for a new dataset suffering from data shortage. However, new target dataset must have some resemblance with the source so that the pre-trained model needs only a slight update. Otherwise, the increased complexity of update procedure renders TL unhelpful.
	

	\subsubsection{Data Mismatch}
	When there is a large mismatch in the data, TL may not be helpful. Alternatively, OL techniques can be used to deal with the mismatch between the training and test datasets. Similar to TL, OL employs a pre-trained DL model but also adapts to the changes in the propagation environment. Furthermore, the main difference between OL and TL is that the former adapts to the same task while the latter learns a different task. Eventually, in OL,  the model updates its parameters in accordance with the new incoming data (Fig.~\ref{fig_TL_OL}). The result is the ability to adapt to the changes in the data due to propagation environment. Thus, OL provides better spectral efficiency than the offline learning techniques, whose performance strongly depends on the learned features in the offline dataset. \textcolor{black}{In~\cite{elbir2019online}, OL was reported to be advantageous over offline training for channel estimation and HB design, where a dedicated CNN is used for each task. Online model update is triggered only if the performance of the first CNN (channel estimation) becomes worse than the least-squares (LS) channel estimate. Then, both CNNs are updated for adapting to the new data.} While several DL-based architectures have been proposed in the literature~\cite{deepCNN_ChannelEstimation,dl_GGui_WCM,Vu2021Jan}, their lack of ability to be adaptive in dynamic channel conditions remains a challenge. In this context, OL is a promising technique for enabling the offline-trained models to adjust to the changing wireless propagation environments.

	\subsection{Learning-Related Challenges}
	\label{sec:Challenges}
	
	\subsubsection{Labeling} {\color{black}In most of the multi-antenna applications, e.g., multi-channel estimation and HB, the DL method requires labeled data, which is unavailable during an online scenario. As a result, the incoming data is first labeled for training and then used for the inference stage.} This procedure would be very inefficient if it requires labeling at every instance of new data. For example, during channel estimation, the input is the received array data and the label is the channel matrix (obtained via an analytical method). Labeling for each incoming data entails obtaining channel matrices analytically, thereby completely precluding any need for DL-based channel estimation. Hence, to leverage the fast prediction performance of DL over analytical approaches, labeling is carried out only if there is a significant mismatch in the data. 	
	
	{\color{black}Alternatively, UL and RL, which do not require labeled datasets, help in adapting to the changes in the propagation environment while optimizing the system performance metrics such as spectral efficiency. For example, UL may be used for beamforming, wherein unlabeled dataset (containing the array steering vector of different directions) trains a neural network by optimizing spectral efficiency~\cite{unsupervisedMag,dl_GGui_WCM}}. Also in RL, the input data need not to be labeled. Instead, award/penalty mechanism, formulated as a function of varying environment characteristics to optimize the objective of the learning problem, is employed. \textcolor{black}{Prior DL works use either MSE, BER, SINR or achievable sum-rate of the overall communication system as the optimizing criterion{\color{black}~\cite{dl_GGui_WCM,elbir2019online}}}. The RL requires longer training times than SL and its performance is usually worse because of the absence of the labels.

	\begin{figure}[t]
		\centering
		{\includegraphics[draft=false,width=.8\columnwidth]{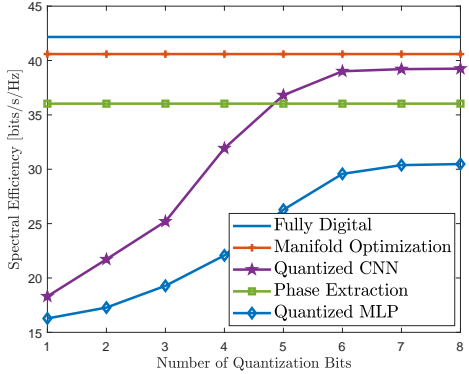} }
		\caption{HB performance of quantized learning models with respect to the number of quantization bits.
		}
		\label{fig_QCNN}
	\end{figure}
	\subsubsection{Learning model complexity}
	Enlarging the width and the depth of the DL model requires a trade-off between the model complexity and learning performance. Recently, to jointly achieve reliable data representation performance and low model complexity, compressed neural networks have gained traction. Quantization of DL model parameters can accelerate training with a slight performance loss~\cite{elbir2019online,mlAtTheEdge}.	 In Fig.~\ref{fig_QCNN}, the HB design performance of the quantized CNN (QCNN) model is presented in terms of spectral efficiency \textcolor{black}{as compared to the state-of-the-art model-based techniques based on manifold optimization (MO)~\cite{elbir2019online} and phase-extraction (PE)~\cite{Vu2021Jan}}. QCNN needs no more than $5$ bits resolution in the learning model parameters to achieve satisfactory performance close to MO, and it provides higher spectral efficiency than the PE-based HB design. We also observe that QCNN performs better than the MLP architecture with quantized parameters, which demonstrates the effectiveness of the convolutional layers.
	
	
	{\color{black}
		\subsubsection{Data and model parallelization}
		One of the main advantage of DL-based methods is lower (up to 40 times) computation times in training and prediction stages using parallel processing techniques, such as GPUs~\cite{mlAtTheEdge}. However, the implementation process for the parallelization on both data and the learning model is a challenging task because of massive number of accesses required to the same data/model. For example, when the dataset is too big to fit into the memory, it is accessed via multiple processors in parallel. Similarly, when the learning model is very large, the model layers are assigned to multiple GPUs to effectively compute the learnable parameters. In order to efficiently process the massive number of data/memory, \cite{parallelizationRef} proposed a tree-search-based algorithm for a $10\times 10$ MIMO detection problem and reduce the processing latency by 29 times with up to approximately 9 times increased energy efficiency. For much larger antenna array setups, further research is required especially when both data and learning model are distributed (e.g. in the FL scenario).}

	\subsection{Communications-Related Challenges}
	The huge processing power required for DL training is usually employed at cloud servers. Pulling training data from individual devices to the cloud server introduces huge overheads that current communication bandwidths do not support. For example, a single long-term evolution frame of $5$ MHz bandwidth and $10$ ms duration carries only $6000$ complex symbols~\cite{elbir2020FL}. Further, a training dataset includes millions of instances of input-output pairs. Thus, data transmission is a bottleneck that should be solved for reliable data transmission and model training.

	\begin{figure}[t]
		\centering
		{\includegraphics[draft=false,width=.8\columnwidth]{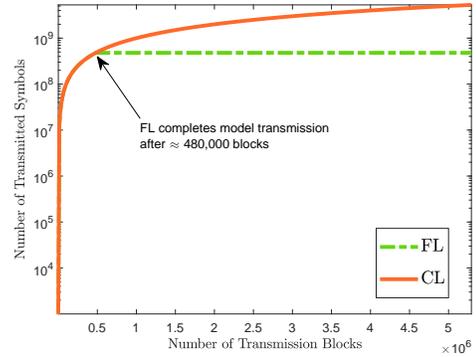} }
		\caption{\color{black}Communication Overhead of CL and FL for HB design problem~\cite{elbir2020FL}. 
		}
		\label{fig_FL}
	\end{figure}
	
	The data congestion is partially addressed by recently proposed FL methods. Instead of transmitting datasets available at the devices, which is called centralized learning (CL), only the gradient information of the model parameters are sent in FL.	It trades-off the sizes of the whole dataset for the number of model parameters. Note that the size of gradients equals the number of learnable model parameters. \textcolor{black}{Since the size of the datasets is large compared to model size in many DL applications~\cite{mlAtTheEdge}}, FL is effective in diminishing the communication overhead. \textcolor{black}{Apart from the gradient-transmission~\cite{elbir2020FL}, there are also model-transmission techniques~\cite{mlAtTheEdge}. However, the former is more energy-efficient~\cite{elbir2020FL}.} The FL-based HB in \cite{elbir2020FL} provides significantly (approximately $15$ times) reduced  data transmission complexity than conventional centralized model training.  Several aspects of FL-based communications require further research, such as scheduling of devices to transmit their gradient data and developing device-level gradient computation. 
	
	\textcolor{black}{Figure~\ref{fig_FL} shows the communication overhead comparison of CL and FL for HB design problem~\cite{elbir2020FL}, wherein $1000$ symbols are assumed to be transmitted in each transmission block. The overhead of CL includes the transmission of the whole dataset to the server whereas FL involves the transmission of model parameters between the users and the server for $100$ communication rounds. We see that FL provides approximately $10$ times lower overhead than CL. }


	{\color{black}
		\section{Summary}
		\label{sec:summ}
		We provided a synopsis of recent developments in learning techniques toward enabling cognitive multi-antenna communications. Here, DL is key to responding to specific challenges imposed by data/hardware complexity, channel dynamics and interference, and power/memory constraints. 
		
		In particular, DL techniques enhance the MSE performance and robustness against the channel dynamics. This is particularly helpful in the highly dynamic channels, such as in mmWave and THz, which employ extremely large arrays. Here, we remark that specific attention is required to address the computational complexity and the related power consumption arising from these higher dimensions.
		
		{\color{black}The learning models are pivoted by a reasonable trade-off between an efficient implementation and satisfactory spectral efficiency performance. Here, a quantized model parameters can be used to reduce model complexity; data/model parallelization is useful for accelerating training times; techniques such as UL and RL are preferred for unlabeled datasets; and FL yields lower communication overhead.}
		
		We conclude that OL and RL are optimal for adaptation to a changing propagation environment; whereas the former requires labeled data leading to longer computation times, the latter processes unlabeled inputs, thus requiring larger training datasets. The combination of both techniques is useful for recent vehicular network applications, wherein the collected datasets are usually unlabeled and environment-dependent.

		When a new array geometry is employed, often the application lacks sufficient training data. In this case, TL delivers the learned features from a pre-trained model and is especially useful for inference from small datasets. 
		
		Futuristic wireless communications require techniques that are privacy-preserving while dealing with several channels and resources in a multi-antenna setup. In this case, the low overhead advantages of FL assume significant importance in decentralized structures such as the upcoming cell-free MIMO communications.
	}

	\bibliographystyle{IEEEtran}
	\bibliography{IEEEabrv,references_overview1}

	\begin{IEEEbiographynophoto} {Ahmet M. Elbir} (S'13-M'16-SM'20) received the  Ph.D. degree from Middle East Technical University  in 2016. He is currently a Research Fellow in University of Luxembourg.
	\end{IEEEbiographynophoto}

	\vspace{-10pt}
	\begin{IEEEbiographynophoto} 
		{Kumar Vijay Mishra} (S'08-M'15-SM'18) obtained a Ph.D. in electrical engineering from The University of Iowa in 2015. He is currently U. S. National Academies Harry Diamond Distinguished Fellow at the United States Army Research Laboratory.
	\end{IEEEbiographynophoto}
\end{document}